\documentstyle[amssymb,epsfig,prb,aps,preprint,tighten]{revtex}

\begin{document}
\title{Slow Crossover in Yb$X$Cu$_{4}$ Intermediate Valence Compounds}
\author{J. M. Lawrence}
\address{University of California, Irvine, CA 92697}
\author{P. S. Riseborough}
\address{Temple University, Philadelphia, PA 19122}
\author{C. H. Booth}
\address{Lawrence Berkeley National Laboratory, Berkeley, CA 94720}
\author{J. L. Sarrao and J. D. Thompson}
\address{Los Alamos National Laboratory, Los Alamos, NM 87545}
\author{R. Osborn}
\address{Argonne National Laboratory, Argonne, IL 60439}
\date{\today}
\maketitle

\begin{abstract}
We compare the results of measurements of the magnetic susceptibility $\chi
(T)$, the linear coefficient of specific heat $\gamma (T)=C(T)/T$ and 4$f$
occupation number $n_{f}(T)$ for the intermediate valence compounds Yb$X$Cu$%
_{4}$ ($X$ = Ag, Cd, In, Mg, Tl, Zn) to the predictions of the Anderson
impurity model, calculated in the non-crossing approximation (NCA). The
crossover from the low temperature Fermi liquid state to the high
temperature local moment state is substantially slower in the compounds than
predicted by the NCA; this corresponds to the ``protracted screening''
recently predicted for the Anderson Lattice. We present results for the
dynamic susceptibility, measured through neutron scattering experiments, to
show that the deviations between theory and experiment are not due to
crystal field effects, and we present x-ray-absorption fine-structure (XAFS)
results that show the local crystal structure around the $X$ atoms is well
ordered, so that the deviations probably do not arise from Kondo Disorder. \
The deviations may correlate with the background conduction electron
density, as predicted for protracted screening.
\end{abstract}




\section{\protect\bigskip Introduction}

The rare earth intermediate valence (IV) compounds\cite{Hewson} are
moderately heavy fermion compounds where the characteristic (Kondo) energy
is large compared to the crystal field splitting ($T_{{\rm K}}>T_{{\rm cf}}$%
). Unlike the truly heavy fermion (HF) compounds\cite{Steglich} (where $T_{%
{\rm K}}<T_{{\rm cf}}$) the IV compounds do not reside close to a quantum
critical point for a transition to antiferromagnetism, so non-Fermi liquid
behavior is neither expected nor observed and the scaling behavior that is
observed does not reflect proximity to such a phase transition.\cite{SCES98}
Furthermore most of the IV compounds are cubic, so that anisotropy and
low-dimensionality are not issues. \ Hence the IV compounds are physical
realizations of the isotropic orbitally degenerate ($N_{J}=2J+1=8$ for Yb)
Anderson Lattice Model\cite{Millis} for three spatial dimensions. This is an
archetypal problem that is both simple and elegant for exploring the physics
of electronic correlations in solids.

A key issue for the Anderson Lattice is the role of lattice coherence, which
can be thought of as dispersive or band-like behavior of the 4$f$ electrons
or alternatively as correlations between the 4$f$ electrons on different
lattice sites. This contrasts the Anderson Lattice (AL) with the Anderson
Impurity Model (AIM) where no such coherence is present. For the HF and IV
compounds the transport behavior, which depends crucially on the periodicity
of the scattering potential, clearly manifests lattice coherence: the low
temperature resistivity (which is finite for the AIM) vanishes for the
periodic compounds, as expected for a system obeying Bloch's Law;\cite
{Hewson,Steglich} the de Haas-van Alphen signals are characteristic of
renormalized $f$ bands;\cite{dHvA} and the optical conductivity exhibits a
Drude response at low temperatures that also reflects renormalized masses. 
\cite{Optical} On the other hand, the dynamic susceptibility $\chi ^{\prime
\prime }(\omega )$ shows a Lorentzian power spectrum with very little $Q$%
-dependence,\cite{LawrenceShapiro} suggesting that the spin/valence
fluctuations are very local and uncorrelated, as expected for the AIM; and
thermodynamic properties that are dominated by the spin/valence fluctuations
such as the temperature dependence of the susceptibility $\chi (T)$,
specific heat $C(T)$ and 4$f$ hole occupation number $n_{f}(T)$ (the Yb
valence is $z$ = 2+$n_{f}$) seem to follow the predictions of the AIM, at
least qualitatively.\cite{Graf,YbL3} Recent theory\cite{Jarrell} of the
Anderson Lattice suggests, however, that there should be observable
differences between the behavior of the AL and the AIM for these quantities.
In particular, ``protracted screening'' can occur in the AL, which means
that the crossover from the low temperature Fermi Liquid state to the high
temperature local moment state is slower for the lattice case than for the
impurity case.

Protracted screening has been invoked\cite{Jarrell2} to explain
photoemission results\cite{Joyce} which show far less temperature dependence
than expected based on the AIM. \ Such results are controversial.\cite
{Reinert} \ To search for such effects in the bulk thermodynamic behavior we
herein compare our measurements\cite{Sarrao} of $\chi (T)$, $C(T)$ and $%
n_{f}(T)$ for the series of related cubic (C15b) compounds Yb$X$Cu$_{4}$ ($X$
= Ag, Cd, In, Mg, Tl and Zn) to the predictions of the Anderson Impurity
Model, calculated within a single approximation scheme, the non-crossing
approximation (NCA). \ By making this comparison for several measurements
for several related compounds, we put strong constraints on the
applicability of the model. \ In addition, we include measurements of
the dynamic susceptibility $\chi ^{\prime \prime }(\omega )$ for $X$ = Ag,
Mg, Tl and Zn; these allow us to determine whether the deviations from AIM
behavior could arise from crystal field effects and also give a fourth
experiment for comparison to the predictions of the model. Finally, we
include measurements of the x-ray absorption fine structure (XAFS) from
the $X$-atom $K$ edges for Yb$X$Cu$_{4}$ and $X$ = Ag, Cd and In and the L$%
_{3}$ edge for $X=$ Tl to determine whether deviations from AIM behavior
could arise from local lattice disorder in the samples.

\section{Experimental details}

The data for the 4{\it f} occupation number $n_{f}(T)$ (Fig. 1), the
susceptibility $\chi (T)$ (Fig. 2) and the linear coefficient of specific
heat $\gamma =C/T$ (Table 1) are taken from Sarrao {\it et al}.\cite{Sarrao}
As explained in that paper, the samples were small, high quality single
crystals grown in $X$Cu flux. \ The specific heat was measured using a
thermal relaxation technique. \ The susceptibility was measured using a
SQUID magnetometer. \ In this work we have subtracted a small ``Curie tail''
(with Curie constant typically of order 10$^{-2}$emu-K/mol) from the
susceptibility. \ The occupation number $n_{f}(T)$ was determined from the
near-edge structure in Yb $L_{3}$ x-ray absorption measurements.

The neutron scattering measurements were performed in the time-of-flight
mode using the LRMECS spectrometer at IPNS (Argonne National Laboratory).
The experimental conditions were similar to those discussed in a recent
study \cite{LawrenceOsborn} of YbInCu$_{4}$; we refer the reader to that
publication for a more detailed account. \ The temperature of the
measurement was 10K, which is considerably smaller than $T_{{\rm K}}$ in
each case. The samples utilized in these experiments were powders, typically
of 50g mass, grown from the melt inside evacuated tantalum tubes for $X$ =
Ag, Mg and Zn; for $X$~=~In and Tl, a large number of single crystals grown
as described in Sarrao {\it et al.}\cite{Sarrao} were powdered to form the
sample. The susceptibilities of these samples were identical to those shown
in Fig.~\ref{chi_fig} except for $X$~=~Ag, where the temperature $T_{max}$
of the maximum in the susceptibility was 10\% smaller for the neutron
sample, and for $X$ = Tl where a fraction of the crystals used to create the
powder had large Curie tails indicative of disorder or impurities. For each
sample, given that the elastic energy resolution (FWHM) varies as $\delta E$ 
$\sim $ 0.07-0.1 {\it E}$_{i}$, we used several incident neutron energies $%
E_{i}$ (e.g. 35, 80 and 150meV) to give greater dynamic range in energy
transfer $\Delta E$ for the measurement. To improve statistics we took
advantage of the lack of $Q$-dependence of the magnetic scattering\cite
{LawrenceShapiro} and grouped detectors into three bins, with average
scattering angle 20$^{\circ }$ (low {\it Q}), 60$^{\circ }$ and 100$^{\circ
} $ (high {\it Q}). \ The scattering was put on an absolute scale by
comparison to the measured scattering of a vanadium sample. \ We also
measured the scattering of YAgCu$_{4}$, LuMgCu$_{4}$, YMgCu$_{4}$, YTlCu$%
_{4} $ and LuZnCu$_{4}$ in order to help determine the nonmagnetic
scattering.

The XAFS experiments were performed on Beam Line 4-3 at the Stanford
Synchrotron Radiation Laboratory (SSRL) using a half-tuned Si(220)
double-crystal monochromator. Samples were ground, passed through a 30 $\mu$%
m sieve and brushed onto scotch tape. Pieces of tape were stacked such that
the change in the absorption at the Yb $L_{{\rm III}}$ edge was $\lesssim$
unity. The samples were the same flux-grown crystals as used in Sarrao {\it %
et al}.\cite{Sarrao} and whose susceptibilities are shown in Fig. \ref
{chi_fig}. For temperature control, we utilized a LHe flow cryostat.

\section{Theoretical details}

The predictions of the Anderson Impurity Model were calculated in the
Non-Crossing Approximation\ (NCA).\cite{BCW} \ This approximation allows for
calculation of all relevant experimental quantities, both static ($\chi $, $%
n_{f}$, $\gamma =C/T$) and dynamic ($\chi ^{\prime \prime }(\Delta E)$), for
realistic orbital degeneracy and for realistic spin-orbit and crystal-field
splitting. \ It agrees well\cite{BCW} over a broad temperature range with
calculations using the Bethe Ansatz, but does produce spurious and
unphysical non-Fermi Liquid artifacts in the low \ frequency density of
states and in the dynamic susceptibility at very low temperatures.\cite{KMH}
These artifacts are not significant for the temperatures and frequencies
studied here. \ The conduction band was assumed to have a Gaussian density
of states of width $W$ and centered at the Fermi energy, i.e. $N(\varepsilon
)=e^{-\varepsilon ^{2}/W^{2}}/(\sqrt{\pi }W)$. \ The hybridization matrix
elements were assumed to be ${\bf k}$-independent, i.e. $V_{kf}=V$, which
may be a good approximation if the thermodynamic and magnetic properties are
predominately determined by the conduction band states within $kT_{\text{K}}$
of the Fermi energy. \ The finite temperature NCA self-consistency equations 
\cite{BCW,Kuramoto} were solved using three overlapping linear meshes. \ The
thermodynamic and spectroscopic results for Ce impurities with spin-orbit
splitting and in various crystal fields were compared with those published
in Bickers {\it et al}\cite{BCW} and good agreement was found despite the
difference in procedure and the different choice of conduction band density
of states.

For Yb IV compounds where crystal fields can be ignored the magnetic orbital
degeneracy is $N_{J}=$ 8 and there are essentially four input parameters for
the AIM calculation: \ the spin-orbit splitting (which we fixed at $\Delta
_{so}=$ 1.3eV, the value typically observed in photoemission experiments\cite
{Joyce}), the width $W$ of the Gaussian conduction band, the hybridization
constant $V$ and the $f$-level energy (relative the Fermi level) $E_{f}$.
Since it is only the conduction states within ${\rm max}\{kT,kT_{{\rm K}}\}$
of the Fermi level that contribute significantly to the experimental
quantities discussed here, the bandwidth parameter $W$ must be chosen to
give the correct value of $N(\epsilon _{\text{F}})$ for the background
conduction band. Assuming the background bandstates in Yb$X$Cu$_{4}$ are
similar to those in Lu$X$Cu$_{4}$, we choose $W$ to reproduce the linear
specific heat coefficient $\gamma ($Lu$)$ of the corresponding Lu$X$Cu$_{4}$
compounds,\cite{Sarrao} using the free electron approximation to convert the
measured value of $\gamma ($Lu$)$ to $N(\epsilon _{\text{F}})$. This is an
important constraint since (as we found in the earlier paper\cite{Sarrao})
it is possible to force fits to $n_{f}(T)$ over the whole temperature range
if unrealistically small values of the bandwidth are used in the model. In Yb%
$X$Cu$_{4}$ compounds, the valence bands have a width of order 10~eV;\cite
{Figueroa,Monachesi} since the 2$\sigma $ full-width of the Gaussian band is 
$\simeq 4W$, the values of $W$ ($\simeq 1eV$, see Table \ref{table1})
obtained from $\gamma ($Lu$)$ are quite reasonable.

Once $\Delta _{{\rm so}}$ and $W$ are fixed, $V$ and $E_{f}$ can be uniquely
determined by fitting to the ground state values of the susceptibility and
occupation number. The values of these parameters are given in Table \ref
{table1}; the Kondo temperature is calculated from the formula 
\begin{equation}
T_{{\rm K}}=(\frac{V^{2}}{\sqrt{\pi }W\left| E_{f}\right| })^{\frac{1}{8}}%
\text{ (}\frac{W}{\Delta _{so}})^{\frac{6}{8}}\text{ }W\text{ }e^{\frac{%
\sqrt{\pi }WE_{f}}{8V^{2}}}\text{ }
\end{equation}
which includes the effect of spin orbit splitting but ignores crystal field
splitting.

For these values of input parameters, we calculated the frequency dependence
of the dynamic susceptibility at 10K and the temperature dependence of the
static susceptibility, the 4f occupation number and the free energy. \ From
the latter we determined the linear coefficient of specific heat by fitting
to the formula $F=E_{0}-\frac{\gamma }{2}(\frac{T}{T_{\text{K}}})^{2}$ in
the temperature range $0.03\leq (T/T_{\text{K}})\leq 0.07$; the values of $%
\gamma $ so determined have a systematic error of order 5-10\%.\qquad

\section{Results and Analysis}

\subsection{Neutron scattering}

The primary goal of the neutron scattering measurements was to determine
whether crystal field excitations can be resolved in these compounds. \
Determination of the magnetic scattering in polycrystals of IV compounds
requires correct subtraction of the nonmagnetic scattering. We have adopted
a variant on the conventional procedure\cite{Goremychkin} for accomplishing
this where the Y$X$Cu$_{4}$ (and/or Lu$X$Cu$_{4}$) nonmagnetic counterpart
to the corresponding Yb$X$Cu$_{4}$ sample is measured to determine the
factor $h(\Delta E)$ = $S$(high $Q$; $\Delta E$;Y)/$S$(low $Q$; $\Delta E;$%
Y) by which the high-$Q$ scattering (which is almost totally nonmagnetic
scattering) scales to the low $Q$ values where the magnetic scattering is
strongest. \ This factor varies smoothly from a value \ $\sim $4-5, at $%
\Delta E=0,$ to a value of 1 at large $\Delta E$ where $Q$-independent
multiple scattering dominates the nonmagnetic scattering. \ (When this
decrease in $h(\Delta E)$ is neglected the nonmagnetic scattering will be
underestimated at large energy transfer.) \ In addition we have included an
energy-independent multiplicative factor $\kappa $ which is needed to
account for the fact that the multiple scattering, which dominates the low-$Q
$ nonmagnetic scattering, has different strengths for Yb than for Y or Lu. \
(When this factor is neglected the magnetic scattering is overestimated.) \
That is, we determine the magnetic scattering from the formula 
\begin{equation}
S_{\text{mag}}\text{(low }Q)=S_{\text{tot}}\text{(low }Q)-\kappa \text{ }%
h(\Delta E)\text{ }S_{\text{tot}}\text{(high }Q\text{)}
\end{equation}
The magnetic scattering is related to the dynamic susceptibility through the
formula 
\begin{equation}
S_{\text{mag}}=\frac{2N}{\pi \mu _{B}^{2}}f^{2}(Q)(1-e^{-\Delta
E/k_{B}T})^{-1}\chi ^{\prime \prime }(Q,\Delta E)
\end{equation}
where $f^{2}(Q)$ is the Yb 4$f$ form factor. \ For each $\kappa $ we fit the
dynamic susceptibility assuming that it is $Q$-independent and assuming a
Lorentzian power spectrum: 
\begin{equation}
\chi ^{\prime \prime }(Q,\Delta E)=\chi _{dc}(T)\text{ }\Delta E\text{ }(%
\frac{\Gamma }{2\pi })\{(\frac{1}{[(\Delta E-E_{0})^{2}+\Gamma ^{2}]})+(%
\frac{1}{[(\Delta E+E_{0})^{2}+\Gamma ^{2}]})\}
\end{equation}
The prediction of the Anderson Impurity Model for $\chi ^{\prime \prime }$
at $T<<T_{{\rm K}}$ also can be fit to this formula; we have included in
Table 1 the values of $E_{0}$ and $\Gamma $ deduced from the fits to the NCA
results. \ The fits to the experimental data include corrections for
absorption and for instrumental resolution. \ The data for every $E_{i}$ is
included in the fit. \ We then find the value of $\kappa $ that gives the
smallest reduced $\chi ^{2}$ for the fit. \ The best-fit values of $\kappa $
were consistently in the range 0.65-0.75; furthermore, the best-fit values
of $\kappa $ gave better agreement (10-20\%) between the fit value of $\chi
_{dc}$ and the value shown in Fig. \ref{chi_fig} than was seen for $\kappa =1
$. \ Use of this multiplicative factor also leads to excellent agreement
between the values of magnetic scattering estimated at different incident
energies. \ In Fig. \ref{neut_fig} we show the data (symbols) for the
magnetic scattering at low $Q$ (fixed average scattering angle $\phi =$20$%
^{\circ }$) and the fits (solid lines). \ (Plots of the data for YbInCu$_{4}$
are given in the earlier publication.\cite{LawrenceOsborn}) \ We note that
in the time-of-flight experiment $Q$ varies with both $E_{i}$ and $\Delta E$
at fixed $\phi $ and consequently the form factor, the absorption and the
instrumental resolution all depend on $E_{i}$; this explains why the data
and the fits for different $E_{i}$ don't overlap. \ The parameters $E_{0}$
and $\Gamma $ for the best fits are given in Table 1, where they are
compared to the values predicted by the AIM calculation. \ We note that our
results agree well with an earlier measurement of YbAgCu$_{4}$.\cite{YbAuCu4}
Although the fits are of reasonable statistical quality with reduced $\chi
^{2}\simeq 1$, the degree of systematic uncertainty due to the assumptions
made about the nonmagnetic scattering is unknown. \ The problem is
especially serious for IV compounds where $T_{{\rm K}}$ is large so that
large incident energies are required, with the result that multiple
scattering is large. \ To be cautious, we expect a 10\% error in the
determination of $E_{0}$ and $\Gamma $. \ 

\subsection{XAFS}

The primary goal of the XAFS measurements was to determine whether $X$/Cu
site interchange is significant in these compounds. Such disorder is a real
possiblity given that YbCu$_{5}$ can grow in the C15b structure,\cite{YbCu5}
and given that UPdCu$_{4}$, which also grows in the C15b struucture, has
been shown to have significant Pd/Cu site interchange.\cite{Booth98b} Our
earlier neutron diffraction measurements\cite{LawrenceKwei} of flux-grown
crystals of YbInCu$_{4}$, coupled with Rietveld refinement, suggest the
average crystal structure is very well ordered. We employ the XAFS technique
here both because the measurement does not depend on lattice periodicity
(i.e. it is a local probe and is thus complimentary to diffraction) and
because it is atomic-species specific. In such a measurement, a
fine-structure function $\chi (k)$ is extracted from the absorption data and
related to the radial bond length distribution around the absorbing $X$
atomic species. This measurement is particularly sensitive to $X$/Cu site
interchange because if any $X$ atoms rest on the Cu sites, a short $X$-Cu
bond length at about 2.55 \AA\ will appear that corresponds to the Cu-Cu
nearest neighbor bond length in the nominal structure. This bond length is
significantly shorter than the nearest neighbor $X$-Cu pairs at $\sim $ 2.93 
\AA\ or the $X$-Yb pairs at $\sim $ 3.06 \AA , and so is easy to resolve if
the site interchange is frequent. In addition to searching for $X$/Cu site
interchange, we also look for more generic disorder in the nominal $X$-Cu
and $X$-Yb pairs by analyzing the distribution widths extracted from the $r$%
-space fits as a function of temperature between 15-300\ K. Bond length
distribution widths should follow a correlated-Debye model\cite{Crozier88}
as a function of temperature if the structure does not have any static
(positional) disorder. This model differs from the usual Debye model because
it includes the correlations in the motions between the atoms in a given
pair. It has been shown to be accurate to within $\sim $5\% for metals.\cite
{Li95b} Any static disorder should manifest as a constant offset, as is
clearly shown, for example, for La$_{1-x}$Ca$_{x}$MnO$_{3}$.\cite{Booth98a}

The XAFS data were fit in $r$-space, following procedures in Ref. [%
\onlinecite{Booth98b,Li95b}]. Initial fits included a short $X$-Cu pair at $%
\sim $2.55 \AA . The relative amplitude of this peak compared to the
amplitude of the nominal $X$-Cu at $\sim $2.93 \AA\ gives the percentage of $%
X$ atoms on Cu sites. Errors in this measurement were determined by assuming
that the fitted statistical $\chi ^{2}$ was equal to the degrees of freedom
in the fit as given by Stern\cite{Stern93} and then increasing the amount of
site interchange until $\chi ^{2}$ increased by one, while allowing all the
other parameters in the fit to vary. These results are presented in Table 
\ref{xafs_table}. For $X=$ In, Cd and Ag the site interchange has the same
value as the error, and hence the results are consistent with the zero site
interchange. \ For YbTlCu$_{4}$ the error is smaller than the site
interchange but it is also smaller than the error in the other cases. \
Furthermore, the fitted Tl-Cu bond length was very low (2.40 \AA ) even
though Tl has the largest atomic radius of all the $X$'s measured. \ We note
that sytematic errors in the fit can arise from use of an incorrect
theoretic backscattering amplitude \cite{Li95b}. \ Given all this, we think
that the error is underestimated for YbTlCu$_{4}$. Therefore, we conclude
that, within $\sim $3\%, none of the materials measured have any $X$/Cu site
interchange.

Once the lack of site interchange was determined, we fit the data assuming
no site interchange and determined the pair distribution widths to search
for any other disorder. Examples of two of these fits are shown in Fig. \ref
{xafs_fig}. The bond lengths of the nominal $X$-Cu and $X$-Yb pairs do not
deviate from the average C15b structure assuming lattice parameters from
Ref. \onlinecite{Sarrao} by more than 0.02 \AA . The distribution widths are
shown in Fig. \ref{debye_fig}, as are the fits to the correlated-Debye
model. \ Parameters of the correlated-Debye fits are given in Table \ref
{xafs_table}. The fits are quite good and the observed values of the static
disorder $\sigma _{stat}^2$ are typical of values seen in well-ordered
compounds.

\subsection{Susceptibility, 4$f$ occupation number and specific heat}

The results of our calculations for $\chi (T)$ and $n_{f}(T)$ for the
different Yb$X$Cu$_{4}$ compounds are shown in Figs. \ref{nf_fig} and \ref
{chi_fig} where they are compared to the experimental behavior.\cite{Sarrao}
The most noteable feature is that for $X$ = Cd, Mg and Zn, the
experimentally determined occupation number $n_{f}(T)$ and the effective
moment $\mu _{eff}^{2}=T\chi /C_{J}$ (where $C_{J}$ is the $J$ = 7/2 Curie
constant) rise towards the high temperature limit more slowly than predicted
by the impurity theory. The effect is somewhat weaker but still apparent for 
$X$ = Ag. For $X$ = Tl the experimental occupation number shows a weak
retardation compared to theory but the experimental susceptibility tracks
the impurity theory up to room temperature. \ The linear coefficients of
specific heat determined from our calculation are given in Table 1 where
they are compared to the experimental values; the experimental and
calculated values of the Wilson Ratio ${\cal R}=(\pi ^{2}R/3C_{J})\chi
(0)/\gamma $ are also compared in the table.

\section{Discussion}

Figures \ref{nf_fig} and \ref{chi_fig} give our basic result: the crossover
from the ground state Fermi Liquid to the high temperature local moment
state is slower for the real materials than predicted by the Anderson
Impurity Model.

There are several reasons why this comparison to theory might be incorrect.
The first is that crystal fields may be large in these compounds. Crystal
fields can cause the ground state multiplet to have a lower degeneracy than
the value $N_{J}=8$ expected for $J=7/2$. Since the AIM results depend on $%
N_{J}$ this could lead to a discrepancy between theory and experiment if $%
N_{J}$ were incorrectly assigned in the calculation. \ \ Indeed, in the
earlier paper\cite{Sarrao} we found that for some cases we could fit the
data assuming a lower degeneracy, e.g. $J=5/2$ for YbMgCu$_{4}$. \ Our
inelastic magnetic neutron scattering experiments were designed to test for
this possibility. \ As mentioned above, there is a systematic error of order
10-20\% in the determination of the peak position and width of the magnetic
scattering. \ For our present purposes, however, the basic point is that the
data can be well-fit by a single broadened Lorentzian whose peak energy $%
E_{0}$ is in reasonable accord with the predicted value (i.e. it is of order 
$T_{{\rm K}}$); there are no sharp crystal field levels in evidence. \ This
also is in accord with the fact that the crystal fields seen in trivalent
compounds such as YbInNi$_{4}$\cite{YbNiCu4} or YbAuCu$_{4}$\cite{YbAuCu4}
are of order 2-4meV which is significantly smaller than $T_{{\rm K}}$ for
most of these compounds.

A second potential problem for our analysis is that if Yb atoms are in a
disordered environment, there can be a distribution of Kondo temperatures.
In this case, the measured susceptibility involves an integral over this
distribution of Kondo temperatures, which would spread the crossover over a
broader range of temperature.\cite{Booth98b,Bernal95} \ This effect may be
relevant in UPdCu$_{4}$, where there is significant ($\approx 25\%$) Pd/Cu
site exchange.\cite{Booth98b} \ \ Our XAFS results for the Yb$X$Cu$_{4}$
compounds\ show that within $\simeq $3\% there is no $X$/Cu site
interchange; furthermore, the observed values of the static disorder $\sigma
_{stat}^{2}$ are typical of values seen in well-ordered compounds. \
Basically, the XAFS results are consistent with a very high degree of order
for the flux-grown crystals. \ We thus believe that Kondo disorder is not a
significant factor for these compounds. (A possible exception is YbZnCu$_{4}$
for which we could not analyse the Zn $K$-edge XAFS because it is
superimposed on the Cu $K$-edge XAFS. Given the chemical similarity of Zn
and Cu, Zn/Cu site disorder seems a real possibility.)

A third problem for this analysis concerns the interpretation of the Yb $%
L_{3}$ x-ray absorption used to determine the occupation number n$_{f}$(T).
\ The analysis\cite{YbL3,Sarrao} relies on rather simple assumptions: the
divalent and trivalent absorption edges are assumed to have the same shape
and the same absorption matrix element, so that n$_{f}$ is determined from
the fractional weights of the two features. \ Similar assumptions are made
for other methods (XPS core levels; valence band UPS; Mossbauer isomer
shift) for determining valence. \ While the resulting systematic error is
not well-understood, it has been argued\cite{Rohler} that the $L_{3}$ method
is superior to use of isomer shift or lattice constant anomalies for the
determination of valence.\ It has been pointed out\cite{Joyce} that the
values of $n_{f}$ estimated from the $L_{3}$ experiments are consistently
and significantly larger than those estimated from valence band
photoemission. \ The latter estimate also requires the assumption of equal
absorption matirix elements for the di- and tri-valent configurations and is
extremely sensitive to choice of background. \ A point in favor of the use
of x-ray absorption is that, unlike photoemission it truly samples the bulk
of the compound; indeed it has been argued\cite{Reinert} that near-surface
reconstruction gives rise to differences between the $L_{3}$ and UPS
estimates of $n_{f}$. \ Whether this is so and what the sytematic errors are
in both measurements are, in our opinion, topics for future research. \ In
any case, the deviations from the AIM that we observe for $n_{f}(T)$ using
the $L_{3}$ measurement (Fig. 1) are qualitatively very similar \ to those
observed for the effective moment (Fig. 2), the measurement of which is not
subject to significant systematic error. (In our earlier paper\cite{Sarrao}
we showed that at room temperature for all cases except $X=$ Ag, $n_{f}$ is
nearly equal to the effective moment.) \ The photoemission analysis yields
an extremely slow crossover, to the extent that the hole occupancy is found
to be $n_{f}\simeq 0.6$ in the high temperature state of YbInCu$_{4}$ which
is much smaller than the value ($>0.9$) that we have observed\cite{Cornelius}
using $L_{3}$. \ The latter value is, in our opinion, more realistic given
the essentially trivalent behavior of the susceptibility of YbInCu$_{4}$ at
high temperature.

We therefore argue that we have ruled out most of the major sources of error
in our analysis, so that we can have a reasonable degree of confidence that
the slow crossover is a real effect. \ As mentioned in the introduction,
recent theory\cite{Jarrell} for the ($S$ = 1/2) Anderson Lattice, treated in
the 1/$d$ approximation (where $d$ is the spatial dimension of the system)
predicts such a slow crossover, referring to it as ''protracted screening''.
\ The degree of slowness is sensitive to the filling of the background
conduction band; the crossover is slowest for partial filling of the
background band. At present it is unclear the extent to which the effect
depends on the orbital degeneracy.

The deviations of the data (Figs. 1 and 2) for Yb$X$Cu$_{4}$ from the
predictions of the AIM are strongest for $X=$ Mg, Zn and Cd, weaker for $X=$
Ag and weakest for $X=$ Tl. \ We note that there is no correlation between
this trend and the magnitude of the Kondo temperature (Table 1) nor of the
ground state occupation number (Fig. 1). \ There {\it does} appear to be a
correlation with the magnitude of the Hall coefficient for Lu$X$Cu$_{4}$: \ $%
R_{0}=-1.8,-1.0$ and $-0.8\times 10^{-10}m^{3}/C$ for $X=$ Mg, Zn, Cd;\cite
{Sarrao} $-0.6\times 10^{-10}m^{3}/C$ for $X=$ Ag;\cite{Cornelius} and $%
<1\times 10^{-11}m^{3}/C$ for $X=$ Tl.\cite{Sarrao} \ This quantity can be
taken as a measure of the conduction electron density in the background
band. \ In a one-band model, where $R_{0}=1/ne$ and where there are 24 atoms
in a unit cell of side 7.1\AA\ these values of $R_{0}$ correspond to the
values 0.52, 0.93, 1.16 and 1.55 electrons per atom for the sequence $\{X=$
Mg, Zn, Cd, Ag$\}$. \ While the one-band assumption is unrealistic,
nevertheless these results suggest that the deviations from the AIM become
weaker as the carrier density becomes larger. \ This accords with theory\cite
{Jarrell} which predicts no deviation for half filling of the background
band and increasing deviation as the conduction electron density decreases.
\ 

We should point out that a slow crossover is expected on rather general
grounds. \ In the single impurity calculations the position of the chemical
potential $\mu $ is set by the host metal conduction band density of states,
and the chemical potential is independent of temperature to order $kT/W$. \
In the hypothetical case of a periodic array of impurities that are
noninteracting and incoherent the appreciable $f$-weight and its asymmetric
temperature dependence will lead to a significant temperature dependence of $%
\mu (T)$ and hence lead to a deviation of $n_{f}(T)$ from the behavior found
in the solution to the single impurity model. \ It is not clear to us the
extent to which this effect contributes to the predicted protracted
screening.

In addition to our main conclusion, we would like to point to two more
features of the analysis. \ The first is that the experimentally determined
linear coefficients $\gamma $ of specific heat for $X$ = Mg, Cd, Ag and Zn
are 20-30\% larger than the values predicted by the AIM. \ This means that
the experimental value of the Wilson ratio ${\cal R}$ is smaller than
predicted by the AIM. This has been noticed before.\cite{Graf,Sarrao} For
YbInCu$_{4}$ there is excellent agreement between the experimental and
calculated values; for YbTlCu$_{4}$ the calculated value of $\gamma $\ is
30\% {\it smaller} than the experimental value. \ The second feature is that
there is considerable disagreement between the measured and predicted values
of the neutron lineshape parameters $E_{0}$ and $\Gamma $ given in Table 1.
We remind the reader that the degree of systematic error for this
determination is poorly understood, due to the difficulty of determining the
nonmagnetic scattering for large $T_{{\rm K}}$ compounds. In addition, all $%
Q $ dependence of the lineshape in polycrystals is automatically averaged
out. In the case of YbInCu$_{4}$, we have confidence in the results, since
an alternate means of subtracting the nonmagnetic scattering\cite
{LawrenceOsborn} agrees with the method used here and since a neutron
scattering study of a single crystal\cite{LawrenceShapiro} showed that the $%
Q $-dependence of the lineshape is quite small. For that case we note that
the experimental peak energy $E_{0}$ agrees well with the prediction of the
AIM but the measured linewidth $\Gamma $ is a factor of two smaller than
predicted, as though the spin fluctations have a longer lifetime than
predicted by the AIM. On the other hand for YbMgCu$_{4}$, YbTlCu$_{4}$ and
YbZnCu$_{4}$ the experimental values of peak energy differ significantly
from the calculated values and the linewidths are {\it larger} than the
predicted values. Such a large linewidth could be a consequence of a $Q$%
-dependence of the peak parameter $E_{0}$($Q$); studies of single crystals
are necessary to establish whether such a $Q$-dependence is present. Finally
we note that if the lineshape parameters for YbAgCu$_{4}$ are increased by
10\% to account for the 10\% smaller value of $T_{max}$ observed for the
neutron sample, they are in excellent agreement with the AIM predictions.

In conclusion, we have given evidence that the crossover from low
temperature Fermi Liquid behavior to high temperature local moment behavior
is slower for periodic intermediate valence compounds than predicted by the
Anderson Impurity Model. \ We have included supporting experiments to rule
out the possibility that the disagreements between the data and the AIM
predictions are due to large crystal fields or to Kondo disorder. \ We feel
that the results are fairly robust. \ It will be interesting to see whether
similar ''protracted screening'' is observed in other compounds, especially
in those based on Ce where the orbital degeneracy is smaller ($N_{J}=6$) and
where the size of the 4$f$ orbital is considerably larger.

\section{Acknowledgments}

Work at UC Irvine was supported by UCDRD funds provided by the University of
California for the conduct of discretionary research by the Los Alamos
National Laboratory. Work at Polytechnic was supported by DOE
FG02ER84-45127. \ Work at Lawrence Berkeley National Laboratory was
supported by the Office of Basic Energy Sciences (OBES), Chemical Sciences
Division of the Department of Energy (DOE), Contract No. DE-AC03-76SF00098.
Work at Los Alamos was performed under the auspices of the DOE. Work at
Argonne was supported by the DOE Office of Science under Contract No.
W-31-109-ENG-38. The XAFS experiments were performed at SSRL, which is
operated by the DOE/OBES.

\newpage \mediumtext
\begin{table}[tbp]
\caption{ Input parameters $W$, $E_{f}$ and $V$ for the AIM calculation; the
calculated Kondo temperature $T_{{\rm K}}$; and the theoretical and
experimental values of the specific heat coefficient $\protect\gamma $, the
Wilson ratio ${\cal R}$ and the neutron lineshape parameters $E_{0}$ and $%
\Gamma $. }
\label{table1}
\begin{tabular}{lllllllll}
Compound & $W$ & $E_{f}$ & $V$ & $T_{{\rm K}}$ & $\gamma $ & ${\cal R}$ & $%
E_{0}$ & $\Gamma $ \\ 
& (eV) & (eV) & (eV) & (K) & $(\frac{mJ}{mol-K^{2}})$ &  & (meV) & (meV) \\ 
\tableline YbTlCu$_{4}$ & 1.286 & -0.50146 & 0.2195 & 514 & 32.1 & 1.26 & 
54.0 & 31.2 \\ 
&  &  &  & Expt: & 24.2 & 1.67 & 38.8 & 32.3 \\ 
YbMgCu$_{4}$ & 1.005 & -0.1897 & 0.128 & 500 & 36.7 & 1.22 & 44.6 & 23.4 \\ 
&  &  &  & Expt: & 53.3 & 0.84 & 31.9 & 34.5 \\ 
YbInCu$_{4}$ & 1$^{1}$ & -0.7442 & 0.232 & 299 & 41.3 & 1.44 & 41.6 & 25.9
\\ 
&  &  &  & Expt: & 41.3$^{1}$ & 1.44$^{1}$ & 41.6 & 13.5 \\ 
YbCdCu$_{4}$ & 0.929 & -0.17576 & 0.098 & 127 & 120.4 & 1.40 &  &  \\ 
&  &  &  & Expt: & 165.6 & 1.02 &  &  \\ 
YbAgCu$_{4}$ & 0.865 & -0.4485 & 0.148 & 95 & 137.9 & 1.29 & 11.3 & 7.1 \\ 
&  &  &  & Expt:: & 198.9 & 0.89 & 9.8$^{2}$ & 6.0$^{2}$ \\ 
YbZnCu$_{4}$ & 1.213 & -0.1984 & 0.1038 & 60 & 296.8 & 1.33 & 5.0 & 4.1 \\ 
&  &  &  & Expt: & 370 & 1.07 & 7.4 & 6.1
\end{tabular}
Notes: \ 1) \ The values of $W$ are deduced from the specific heat of the
corresponding Lu$X$Cu$_{4}$ compound and the experimental specific heat
coefficients have been corrected for the non-4$f$ contribution by
subtracting the specific heat coefficient of the Lu$X$Cu$_{4}$ compound. \ \
Since the groundstate of LuInCu$_{4}$ is semimetallic\cite{Figueroa} and
does not serve for these estimates, we have arbitrarily set $W$ = 1eV and we
have subtracted a corresponding amount (8.7mJ/molK$^{2})$ from the specific
heat coefficient. \ 2) \ For comparison to theory these values should be
increased by 10\% to account for the fact that the neutron scattering sample
had a Kondo temperature that is 10\% smaller than the sample used to measure 
$\chi $ and $n_{f}$.
\end{table}

\mediumtext
\begin{table}[tbp]
\caption{XAFS fit results between 2.0 and 3.2 \AA. Fits only include the
nominal structural peaks $X$-Cu at $\sim$2.9 \AA\ and $X$-Yb at $\sim$3.06
\AA. $S_0^2$ is an amplitude reduction fact that sets the scale for the
distribution variance measurements. See text for details. }
\label{xafs_table}
\begin{tabular}{ccccccc}
&  & \multicolumn{2}{c}{$\Theta_{{\rm cD}}$ (K)} & \multicolumn{2}{c}{$%
\sigma^2_{{\rm stat}}$ (\AA$^2$)} & $X$/Cu \\ 
$X$ & $S_0^2$ & Cu & Yb & Cu & Yb & interchange \\ 
\tableline Tl & 0.89(5) & 230(5) & 230(5) & 0.0005(4) & 0.0005(5) & 4(1)\%
\\ 
In & 1.04(5) & 252(5) & 280(5) & 0.0009(4) & 0.0011(5) & 2(3)\% \\ 
Cd & 0.98(5) & 240(5) & 255(5) & 0.0007(4) & 0.0010(5) & 5(5)\% \\ 
Ag & 0.91(5) & 250(5) & 235(5) & 0.0008(4) & 0.0006(5) & 2(2)\%
\end{tabular}
\end{table}
\widetext

\begin{figure}[tbp]
\caption{ The 4$f$ hole occupation number $n_{f}(T)$ versus temperature for
the Yb$X$Cu$_{4}$ compounds. \ The open circles are the experimental data
and the solid lines are the predictions of the Anderson Impurity Model
(AIM), with input parameters given in Table 1. }
\label{nf_fig}
\end{figure}

\begin{figure}[tbp]
\caption{ The magnetic susceptibility $\protect\chi $(T) (open circles) and
the effective moment $\protect\mu _{eff}^{2}=T\protect\chi /C_{J}$ (solid
circles) for the Yb$X$Cu$_{4}$ compounds. The solid (dotted) lines are the
prediction of the AIM for the susceptibility (effective moment). }
\label{chi_fig}
\end{figure}

\begin{figure}[tbp]
\caption{ The inelastic magnetic neutron scattering spectra for the Yb$X$Cu$%
_{4}$ compounds at $T$ = 10~K. \ For each sample, two or three values of the
incident energy $E_{i}$ were used to increase the dynamic range. \ (YbMgCu$%
_{4}$: solid circles, $E_{i}$ = 150~meV; open circles, $E_{i}$ = 80meV; open
diamonds, $E_{i}$ = 35meV. \ YbTlCu$_{4}$: solid circles, $E_{i}$ = 150meV;
open circles, $E_{i}$ = 80meV. \ YbZnCu$_{4}$: solid circles, $E_{i}$ =
80meV; open circles, $E_{i}$ = 25meV; open diamonds, $E_{i}$ = 15meV. \
YbAgCu$_{4}$: solid circles, $E_{i}$ = 60meV; open circles, $E_{i}$ =
25meV.) \ The solid lines represent fits to the Lorentzian power spectrum
(see text) with the parameters given in Table 1. }
\label{neut_fig}
\end{figure}

\begin{figure}[tbp]
\caption{XAFS data and fits on (a) YbAgCu$_4$ and (b) YbCdCu$_4$. The outer
envelope shows $\pm$ the amplitude (or modulus) and the oscillating inner
curve is the real part of the complex Fourier transform of $k^3\protect\chi%
(k)$. The $r$-axis includes different phase shifts for each coordination
shell, so that the $r$ of a given peak appears somewhat longer than the
actual pair distance $R$. These transforms are from 2.5-15 \AA$^{1}$ and
Gaussian narrowed by 0.3 \AA$^{-1}$. Fits are from 2.0 to 3.2 \AA. }
\label{xafs_fig}
\end{figure}

\begin{figure}[tbp]
\caption{ The variance in (a) the $X$-Cu pair distribution at $\sim$2.9 \AA,
and (b) the $X$-Yb pair distribution at $\sim$ 3.06 \AA. The temperature
dependence follows a correlated-Debye model well for all the data, as shown.
Most of the low temperature data are very similar, except the Tl-Cu and
Tl-Yb $\protect\sigma^2$'s are somwhat lower due to the greater mass of Tl
compared to the other $X$'s. }
\label{debye_fig}
\end{figure}

\end{document}